\def\i{{\rm i}}

\def\e{{\rm e}}

\def\binom#1#2{\left(\begin{array}{@{}c@{}} #1 \\ #2 \end{array}\right)}
\newtheorem{conj}{Conjecture}
\DeclareFontShape{OML}{cmm}{m}{b}{%
   <-> cmmib10}{}
\DeclareMathAlphabet{\mathbf}{OML}{cmm}{m}{b}
\DeclareSymbolFont{boldletters}{OML}{cmm}{m}{b}
\DeclareMathSymbol{\balpha}{\mathord}{boldletters}{11}
\DeclareMathSymbol{\bbeta}{\mathord}{boldletters}{12}
\DeclareMathSymbol{\bgamma}{\mathord}{boldletters}{13}
\DeclareMathSymbol{\bomega}{\mathord}{boldletters}{33}
\DeclareMathSymbol{\bsigma}{\mathord}{boldletters}{27}
\DeclareMathSymbol{\btau}{\mathord}{boldletters}{28}
\documentstyle[12pt,amssymb]{article}
\begin{document}
\title{\large
\bf The quantum symmetric XXZ chain at $\Delta=-\frac{1}{2}$,\\
alternating sign matrices and plane partitions}
\author{M.T. Batchelor and J. de Gier\\
{\em Department of Mathematics, School of Mathematical Sciences,}\\
{\em Australian National University, Canberra ACT 0200, Australia}
\and
B. Nienhuis\\
{\em Instituut voor Theoretische Fysica, Universiteit van Amsterdam,}\\
{\em 1018 XE Amsterdam, The Netherlands}}
\date{\today}
\maketitle 
\begin{abstract}
We consider the groundstate wavefunction of the quantum symmetric 
antiferromagnetic XXZ chain
with open and twisted boundary conditions at $\Delta=-\frac{1}{2}$,
along with the groundstate wavefunction of the corresponding O($n$) 
loop model at $n=1$. 
Based on exact results for finite-size systems, sums involving 
the wavefunction components, and in some cases the largest component
itself, are conjectured to be directly related to
the total number of alternating sign matrices 
and plane partitions in certain symmetry classes.  
\end{abstract}

\vskip 5mm

Very recently Razumov and Stroganov \cite{RS} have made some remarkable conjectures 
in which the number of $n \times n$ alternating sign matrices appear in certain 
properties of the ground state 
wavefunction of the antiferromagnetic XXZ Heisenberg chain 
at $\Delta=-\frac{1}{2}$ defined on an odd number of sites $L$.
At this point, the ground state energy of the XXZ hamiltonian 
\begin{equation}
H = - \frac{1}{2} \sum_{j=1}^{L} \left( \sigma_j^x \sigma_{j+1}^x + \sigma_j^y
\sigma_{j+1}^y + \Delta \sigma_j^z \sigma_{j+1}^z \right) 
\label{eq:XXZham}
\end{equation}
with periodic boundary conditions, is simply \cite{ABB,Str}
\begin{equation}
E_0 = - 3 L/4.
\label{eq:triv}
\end{equation}
The groundstate is 2-fold degenerate with spin $S^z = \pm \frac12$.
Define the $p$-sum $S_L^{(p)}$ of the components of the 
groundstate wavefunction $\psi_0$ by
\begin{equation}
S_L^{(p)} = \sum_{s=1}^{\scriptsize \binom{L}{\left[ L/2 \right]}} 
\psi_0(s)^p
\label{psums}
\end{equation}
where ${\left[ L/2 \right]}$ defines the integer part of $L/2$
and ${\scriptsize \binom{m}{n}}$ is the usual Binomial coefficient.
Normalise the groundstate wavefunction such that the smallest 
nonzero entry is unity.
Then based on explorations on finite systems of size $L=2n+1$,
Razumov and Stroganov have made the following conjectures (among others)

\vskip 2mm
\noindent
{\bf Conjecture} (Razumov and Stroganov \cite{RS}) 
{\em The largest element of the groundstate
wavefunction is given by $A_n$, where
\begin{equation}
A_n = \prod_{j=0}^{n-1} \frac{(3j+1)!}{(n+j)!}.
\end{equation}
}

\vskip 2mm
\noindent
{\bf Conjecture} (Razumov and Stroganov \cite{RS}) 
{\em The 1- and 2-sums for the groundstate wavefunction are given by  
\begin{eqnarray}
S_{2n+1}^{(1)} &=& ( \sqrt 3 )^n {\cal N}_n \nonumber  \\
S_{2n+1}^{(2)} &=& {\cal N}_n^2 \nonumber 
\end{eqnarray}
where
\begin{equation}
{\cal N}_n = \left( \frac{\sqrt 3}{2} \right)^n \frac{2 \cdot 5 \cdots (3n-1)}
                                      {1 \cdot 3 \cdots (2n-1)} A_n .
\end{equation}
}
 
The numbers $A_n$ form the well known sequence
$1, 2, 7, 42, 429, 7436, 218348, \ldots$
of $n \times n$ alternating sign matrices \cite{MRR},
or equivalently, the total number of descending plane partitions
with largest part less than or equal to $n$ (see, e.g., \cite{B} and references
therein).
The 2-sum $S_{L}^{(2)}$ gives the squared norm of the groundstate wavefunction.

Two variations of the $XXZ$ chain with such a special 
ground state energy at $\Delta = - \frac{1}{2}$ are also known.
Consider first the hamiltonian (\ref{eq:XXZham}), with twisted boundary conditions
\begin{equation}
\sigma_{L+1}^z = \sigma_1^z, \qquad \sigma_{L+1}^{\pm} = \e^{\i \phi} \sigma_1^{\pm}   
\label{twisted}
\end{equation}
where $\sigma^{\pm} = \sigma^x \pm \i \sigma^y$ are the standard raising and lowering
operators.
For $\Delta = - \cos \lambda$ and twist angle $\phi = 2 \lambda$, with $\lambda =
\pi/3$ the ground state energy is again given by (\ref{eq:triv}), 
this time for $L$ even \cite{ABB}.
In this case the groundstate wavefunction is complex.
Nevertheless, we also find some surprising connections with the numbers $A_n$. 
We choose the normalisation such that the smallest elements have amplitude
1 and the wavefunction is invariant under the combined transformation
of interchanging left and right and complex conjugation.
Again, based only on exploration of $\psi_0$ for finite sizes, 
there are patterns to be observed.
Our conjectures are

\begin{conj}
The 1-sum of the groundstate wavefunction of the twisted XXZ chain at
$\Delta = - \frac{1}{2}$ is given by
$$
S_{2n}^{(1)} = 3^{n/2} A_n .
$$
\end{conj}
\begin{conj}
The 2-sum of the groundstate wavefunction of the twisted XXZ chain at
$\Delta = - \frac{1}{2}$ is given by
$$
S_{2n}^{(2)} = A_n^2 .
$$
\end{conj}

Now consider the XXZ chain with the open boundary conditions \cite{ABB,ABBBQ}  
\begin{equation}
H = - \frac{1}{2} \left[ \sum_{j=1}^{L-1} \left( \sigma_j^x \sigma_{j+1}^x + 
\sigma_j^y \sigma_{j+1}^y + \Delta \sigma_j^z \sigma_{j+1}^z \right) - 
\i \sin\lambda (\sigma_1^z - \sigma_L^z) \right]. 
\label{eq:XXZopen}
\end{equation}
This hamiltonian maps directly to that of the $Q$-state Potts chain, with
$\sqrt Q = 2 \cos \lambda$.
It is also quantum $U_q[sl(2)]$-invariant \cite{PS}.
In the previous case the twisted boundary conditions (\ref{twisted}) 
ensure that the groundstate of the XXZ chain maps to that of the 
periodic Potts model.
At $\Delta=-\frac12$ the groundstate energy of (\ref{eq:XXZopen}) is given by
$E_0 = -3(L-1)/4$ for both $L$ even and odd \cite{ABBBQ} (see also \cite{FSZ}).
The groundstate wavefunction is again complex.
We again
choose the normalisation such that the smallest elements have amplitude
1 and the wavefunction is invariant under the combined transformation
of interchanging left and right and complex conjugation.
For this case we are led to the following conjectures.\footnote{In this case
we found Sloane's On-Line Encyclopaedia of Integer Sequences to be most helpful,
see {\tt http://www.research.att.com/}$\sim${\tt njas/sequences/}.} 

\begin{conj}
The 1-sum of the groundstate wavefunction of the quantum invariant XXZ chain at
$\Delta = - \frac{1}{2}$ is given by
\begin{eqnarray}
S_{2n}^{(1)} &=& 3^{n/2} A_{\rm v} (2n+1) \nonumber \\
S_{2n-1}^{(1)} &=& 3^{(n-1)/2} N_8(2n) \nonumber
\end{eqnarray}
where $A_{\rm v} (2n+1)$ is the number of $(2n+1) \times (2n+1)$
vertically symmetric alternating sign matrices (\cite{K}, Theorem 2) given by,
\begin{equation}
 A_{\rm v} (2n+1) = (-3)^{n^2} \prod_{i=1}^{2n+1} \prod_{j=1}^{n}
\frac{3(2j-i)+1}{2j-i+2n+1}
\end{equation}
and $N_8(2n)$ is the number of cyclically symmetric transpose
complement plane partitions \cite{B,MRR2} given by
\begin{equation}
N_8(2n) = \prod_{i=1}^{n-1} (3i+1) \frac{(2i)!(6i)!}{(4i)!(4i+1)!}.
\end{equation}
\end{conj}
\begin{conj}
The 2-sum of the groundstate wavefunction of the quantum invariant XXZ chain at
$\Delta = - \frac{1}{2}$ is given by
\begin{eqnarray}
S_{2n}^{(2)} &=& A_{\rm v} (2n+1)^2 \nonumber \\
S_{2n-1}^{(2)} &=& N_8(2n)^2 . \nonumber
\end{eqnarray}
\end{conj}

Several comments are in order.
In the past, wavefunctions in Bethe Ansatz systems such as the XXZ chain
have been traditionally ignored because of their unwieldly nature.
Combined with Razumov and Stroganov's observations for the periodic odd chain,
we now see that there is some remarkable structure in the first two $p$-sums
for the groundstate wavefunction of the antiferromagnetic XXZ chain at 
$\Delta=-\frac{1}{2}$.
Presumably the conjectured results can be proved in each case via the
Bethe Ansatz form of the wavefunctions, which {\em are} known. 
We hope that our findings will spark further interest in their properties.

Even more remarkable is that the observed structure is related to alternating 
sign matrices, or rather plane partitions.
For the two cases considered here the trivial groundstate energy can be viewed
as a result of the trivial representation in the related Temperley-Lieb-Jones 
algebra \cite{ABB}.   
That may also have a bearing on the wavefunction.
As noted above the wavefunctions are complex.
However, they are real in the corresponding dense O($n$) loop model \cite{BKW,BN}, 
in which closed loops carry fugacity $n = 2 \cos\lambda$ (in this case $n=1$).
The numbers appearing in the first few groundstate wavefunctions for open and 
periodic boundary conditions are given in Tables 1-3.
In each case we normalise the wavefunctions such that the smallest element 
is unity.\footnote{Note that for the loop models, the $p$-sums defined 
in (\ref{psums}) are over sums smaller than their binomial counterparts in the
XXZ chain.} 
Four further conjectures suggest themselves. 

\begin{table}[h]
\begin{center}
\begin{tabular}{||c|c|l|l|r||}
\hline
$L$  & $n$ & $\psi_0$  & {\rm multiplicity} & $S_{2n}^{(1)}$    \\ \hline
  2  &1    & (1) & (1)   & 1    \\
  4  &2    & (2,1) & (1,1)   & 3    \\
  6  &3    & (11,5,4,1) & (1,2,1,1)   & 26    \\
  8  &4    & (170,75,71,56,50,30,14,6,1) & (1,2,1,2,1,1,4,1,1)   & 646   \\ \hline
\end{tabular}
\vspace{0.4cm}
\caption{Groundstate wavefunctions of the O(1) loop model with open 
boundaries ($L$ even). Note that by $\psi_0=(11,5,4,1)$ with multiplicity $(1,2,1,1)$
we mean $\psi_0=(11,5,5,4,1)$ etc.}
\end{center}
\end{table}

\begin{table}[h]
\begin{center}
\begin{tabular}{||c|c|l|l|r||}
\hline
$L$  & $n$ & $\psi_0$  & {\rm multiplicity} & $S_{2n-1}^{(1)}$     \\ \hline
  1  &1    & (1)   & (1) & 1    \\
  3  &2    & (1)  & (2) & 2    \\
  5  &3    & (3,1)  & (3,2) & 11   \\
  7  &4    & (26,10,9,8,5,1)  & (4,2,2,2,2,2)& 170   \\ \hline
\end{tabular}
\vspace{0.4cm}
\caption{Groundstate wavefunctions of the O(1) loop model with open boundaries ($L$ odd).}
\end{center}
\end{table}

\begin{table}[h]
\begin{center}
\begin{tabular}{||r|c|l|l|r||}
\hline
$L$  & $n$ & $\psi_0$  & {\rm multiplicity} & $S_{2n}^{(1)}$    \\ \hline
  2  &1  & (1)   & (1) & 1    \\
  4  &2  & (1)   & (2) & 2    \\
  6  &3  & (2,1) & (2,3) & 7    \\
  8  &4  & (7,3,1)& (2,8,4)   & 42   \\ 
 10  &5  & (42,17,14,6,4,1) & (2,10,5,10,10,5)   & 429   \\ \hline
\end{tabular}
\vspace{0.4cm}
\caption{Groundstate wavefunctions of the O(1) loop model with periodic boundaries.}
\end{center}
\end{table}

\begin{conj}
For open boundary conditions, the largest element of the O(1) loop model
wavefunction is given by $N_8(2n)$ for $L=2n$ and $A_{\rm v}(2n-1)$ for $L=2n-1$.
\end{conj}

\begin{conj}
The 1-sum of the groundstate wavefunction of the O(1) loop model with 
open boundary conditions is given by
\begin{eqnarray}
S_{2n}^{(1)} &=& A_{\rm v} (2n+1) \nonumber \\
S_{2n-1}^{(1)} &=& N_8(2n). \nonumber
\end{eqnarray}
\end{conj}

\begin{conj}
For periodic boundary conditions and $L=2n$, the largest element of the O(1) loop model
wavefunction is given by $A_{n-1}$.
\end{conj}

\begin{conj}
The 1-sum of the groundstate wavefunction of the O(1) loop model with
periodic boundary conditions and $L=2n$ is given by
$S_{2n}^{(1)} = A_n$.
\end{conj}

\begin{table}[h]
\begin{center}
\begin{tabular}{||c|c|l|l|r||}
\hline
$L$  & $n$ & $\psi_0$  & {\rm multiplicity} & $S_{2n-1}^{(1)}$     \\ \hline
  1  &1    & (1)    & (1) & 1    \\
  3  &2    & (1)    & (1) & 3    \\
  5  &3    & (4,1)  & (1,1) & 25    \\
  7  &4    & (49,14,6,1)  & (1,2,1,1) & 588    \\ 
  9  &5    & (1764,567,525,266,150,132,49,27,8,1) & 
(1,1,2,2,1,1,2,2,1,1) & 39204    \\ \hline 
\end{tabular}
\vspace{0.4cm}
\caption{Groundstate wavefunctions of the O(1) loop model with periodic 
boundaries ($L$ odd). The multiplicities are divided by $L$.}
\end{center}
\end{table}

In view of the relative simplicity of the loop model results, we also considered
the loop version of the XXZ chain at $\Delta=-\frac{1}{2}$ for odd sites.
The numbers appearing in the first few groundstate wavefunctions 
are given in Table 4.
Again we have normalised the wavefunctions such that the smallest element is unity.
Two further conjectures are
\begin{conj}
For periodic boundary conditions and $L=2n-1$, the largest element of the 
O(1) loop model wavefunction is given by $A_{n-1}^2$.
\end{conj}
\begin{conj}
The 1-sum of the groundstate wavefunction of the O(1) loop model with
periodic boundary conditions and $L=2n-1$ is given by
$S_{2n-1}^{(1)} = {\cal N}_n^2$.
\end{conj}
 
These are to be compared with those of Razumov and Stroganov \cite{RS} given above.
An interesting comparison can be made between Conjectures 7 \& 9 for
periodic boundary conditions with even and odd system sizes.
On the one hand, for $L$ even, the largest element in the wavefunction is given by 
$A_{n-1}$, which we recall is also equivalent to the total number of descending 
plane partitions with largest part less than or equal to $n-1$.
This quantity is also the number of totally symmetric 
self-complementary plane partitions in ${\cal B}(2n-2,2n-2,2n-2)$, 
a cubic box of linear size $2n-2$.
On the other hand, for $L$ odd, the largest element is given by $A_{n-1}^2$.
However, $A_{n-1}^2$ is the number 
of cyclically symmetric self-complementary
plane partitions in ${\cal B}(2n-2,2n-2,2n-2)$ \cite{B}.
Further, Conjecture 5 states that for $L$ even and open boundaries,
the largest element in the wavefunction is given by the
number of cyclically symmetric transpose complement plane partions
in ${\cal B}(2n,2n,2n)$.
Such results may be a hint that the underlying relations are more directly
with plane partitions rather than alternating sign matrices.

The appearance of alternating sign matrices may also be a remnant of the
eigenspectrum symmetry between the points $\Delta=-\frac{1}{2}$ and
$\Delta=\frac{1}{2}$.
Afterall, the point $\Delta=\frac{1}{2}$ corresponds to the ice model,
which has a well documented connection with alternating sign matrices
with domain wall
boundary conditions in the vertex formulation \cite{B}.
In particular, Kuperberg \cite{K} has shown how to derive a number of results
for various symmetry classes of alternating sign matrices
from the six-vertex model with different
boundary conditions.
We find that to be a very interesting paper.

\vskip 5mm
This work has been supported by the Australian Research Council (ARC) 
and by the `Stichting voor Fundamenteel
Onderzoek der Materie (FOM)',  which is financially supported by the
`Nederlandse Organisatie voor Wetenschappelijk Onderzoek (NWO)'.


\end{document}